\documentstyle[psfig]{lamuphys}
\makeatletter
\let\chapter\hid@chapter
\makeatother

\begin{document}
\pagenumbering{arabic}
\titlerunning{ The SBF and GC Population of M87}
\title{The Surface Brightness Fluctuations and Globular Cluster Populations
of M87 and its Companions}

\author{Eric H. Neilsen\, Jr., 
        Zlatan I. Tsvetanov,
        Holland C. Ford}

\institute{Johns Hopkins University, Baltimore, MD, USA}

\maketitle


\begin{abstract}
Using the surface brightness fluctuations in HST WFPC-2 images, we
determine that M87, NGC~4486B, and NGC~4478 are all at a distance of
$\sim 16$ Mpc, while NGC~4476 lies in the background at $\sim 21$ Mpc.
We also examine the globular clusters of M87 using archived HST
fields. We detect the bimodal color distribution, and find that the
amplitude of the red peak relative to the blue peak is greatest near
the center. This feature is in good agreement with the merger model of
elliptical galaxy formation, where some of the clusters originated in
progenitor galaxies while other formed during mergers.
\end{abstract}
%
\section{Introduction}
%

An accurate estimate of the distance to M87 is important for the study
of its other properties, in particular for determining the correct
physical scales and luminosities from measured angular separations and
aparent magnitudes. As the brightest, central galaxy in its portion of
the Virgo cluster, M87 has a rich environment, including several
companions and exeptionally large population of globular clusters (GC).
By measuring accurate distances we can separate true neighbors from
close projections, while mapping global properties of the GC population
can give important clues for the merger history of M87 and the Virgo
cluster in general.

\index{SBF} \index{surface brightness fluctuations} 
Because of its smooth morphology, the surface brightness fluctuation
(SBF) method of distance determination is ideally suited for
application to M87. The method works by measuring the statistical
effects of the galaxy being made up of a large number individual
stars.

The basis for the method is clearest if one considers the idealized
case of a galaxy with uniform surface brightness, whose stars all have
the same luminosity, $l$. In an image of this galaxy, each pixel will
contain the light from some number of stars, $n(x,y)$. The number of
stars per pixel will have an average $\bar{n}$, and a standard
deviation $\sqrt{\bar{n}}$. By examining the image, one can measure
the mean flux in a pixel, $\bar{f}=\bar{n}l$, and its standard
deviation, $\sigma_{f}=\sqrt{\bar{n}}l$. Finally, one can use these to
calculate the flux of a star in the galaxy:
$l=\sigma_{f}^{2}/\bar{f}$. If we can estimate the absolute luminosity
from this star, it may be used as a ``standard candle,'' and the
distance can be measured.

In practice, there are a number of complications. The morphology of
the galaxy is not uniform, the variance due to stellar statistics must
be separated from variance due to noise and contaminating objects, and
galaxies are made up of stars with a variety of luminosities. All of
these difficulties can be overcome to a large extent. A complete
description of the process of SBF measurement can be found in the
original paper by
\cite{tonry1}, or in \cite{neil1}. 

\index{globular clusters!luminosity function}
A second method of distance determination uses the globular cluster
luminosity function (GCLF). Assuming that the luminosity function of
globular clusters can be estimated, one may use it as a standard
candle. It is particularly convenient to measure the GCLF on data for
which one has already measured SBF, because it is necessary to locate
the globular clusters in the image to obtain an SBF measurement. The
creation of a catalog of globular clusters also allows for the study
of the spatial and color distributions of the globular cluster
population.
\vspace{-3 mm}
\section{The Data}
\begin{figure}
\centerline{\psfig{figure=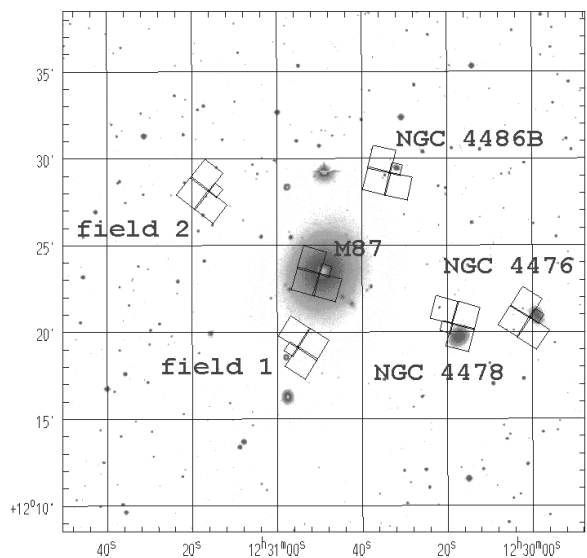,width=3.0in}}
\caption{The placement of the fields around M87. The image on which
our fields are shown was created using an image from the GASP plates and
an image taken by Hintzen et al (1993), supplied courtesy the NCSA
Astronomy Digital Image Library.
\label{fig:place} }
\vspace{-5 mm}
\end{figure}

\vspace{-2 mm}
The Hubble Space Telescope archive provided all of the images used in
this study. Figure~\ref{fig:place} shows the placement of the fields
around M87.
For each field, we used HST WFPC-2
images with a filter approximating the $I$ band (F814W) and a filter
approximating the $V$ band (either F555W or
F606W). Table~\ref{tab:data} presents, for each field, the projected
separation from the center of M87 to the center of the field, the
filters used, the total exposure times, and the number of exposures in
each filter.
\begin{table}
\centerline{
\begin{tabular}{llllll} \hline \hline
Field &  Sep. (arcmin.) & Bands $\;$& Filters & $t_{exp}$ (sec.) & \# exp. \\
 \hline
M87 center & 0.4    & V, I  & F555W, F814W $\;$ & 2400, 2400 & 4, 4  \\
field 1    & 4.5    & V, I  & F606W, F814W  & 3380, 4200 & 5, 4  \\
NGC~4486B  & 6.7    & V, I  & F555W, F814W  & 1800, 2000 & 3, 4  \\
field 2    & 8.2    & V, I  & F606W, F814W  & 1800, 13400 & 3, 5  \\
NGC~4478   & 8.2    & V, I  & F606W, F814W  & 16800, 16500 $\;$ & 6, 6  \\
NGC~4476   & 12.0   & V, I  & F555W, F814W  & 2400, 2400 & 4, 4  \\ \hline
\end{tabular}
}
\caption{The data used in this study. \label{tab:data}} 
\vspace{-5 mm}
\end{table}

\vspace{-3 mm}
\section{The Distances}
\vspace{-2 mm}
\index{distance} \index{NGC~4486B} \index{NGC~4476} \index{NGC~4478}
In table~\ref{tab:distances}, we present the distances determined
through our SBF and GCLF measurements. We also report the distance as
determined by the planetary nebula luminosity function, as given by
\cite{ciar1}. 
\vspace{-1 mm}
\begin{table}
\centerline{
\begin{tabular}{lllll} \hline \hline
Method  & M87         & NGC~4486B $\;$& NGC~4478 $\;$& NGC~4476 \\ \hline
SBF     & $15.8 \pm 1.0 \;$ & $16.2 \pm 1.0 \;$ & $15.1 \pm 1.0 \;$ & $21.1 \pm 1.1$ \\
GCLF    & $14.2 \pm 1.7$ &                 & $16.4 \pm 3.7$ & $19.2 \pm 3.4$ \\
PNLF    & $14.9 \pm 0.7$ & & & \\ \hline
\end{tabular}
}
\caption{The distances to M87 and its companions in Mpc, by three methods.
\label{tab:distances}}
\vspace{-5 mm}
\end{table}

To calculate the SBF distances, we measured the mean color in the
utilized region of the galaxy using the two filters and converted to
$V-I$ using the calibration of \cite{holtz}, and the F814W SBF
calibration of \cite{ajhar}. To calculate the GCLF distances, we
estimate the peak of the GCLF using \cite{ashman}, whose theoretical
results agree well with the observational calibration of \cite{whit}.
The distances measured by the different methods agree well. From these
distances, it appears that NGC~4486B and NGC~4478 are genuine
companions of M87, while NGC~4476 lies in the background. 
\vspace{-3 mm}
\section{The Globular Cluster System of M87}
\vspace{-2 mm}
\index{globular clusters} \index{globular clusters!colors}
In four of our fields, (the center of M87, field~1, the NGC~4486B
field, and field~2) we expect our globular cluster candidate catalog
to be dominated by clusters from M87; this allows us to study the
color distribution in several locations. It has been known for some
time that the cluster population becomes bluer with distance from the
center (see Strom et al. 1981), and a variety of explanations have been
proposed. \cite{strom} suggests that the clusters all formed at the
same time, probably preceding the formation of the galaxy itself, and
that the gradient in colors has the same origin as the gradient in the
color of the integrated halo. In contrast, \cite{ashman2} claim that
M87 was formed as the result of mergers, which play a critical role in
the formation of the globular cluster population. In this model,
during mergers the red clusters are formed in the center of M87 from
the gases in the merging galaxies, while the blue clusters,
originally associated with the progenators, form the more extended
cluster halo.

\begin{figure}
\centerline{
\psfig{figure=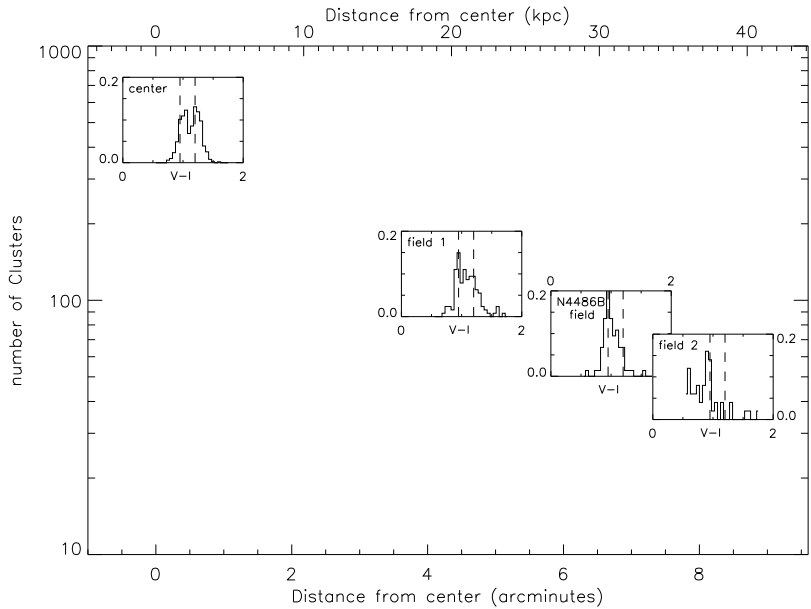,width=4.5in}
}
\caption{The color distribution of M87's globular cluster in several
fields. The placement of each plot represents the distance from the
center of the field to the center of M87, and the total number of
clusters with good $V-I$ colors data. The dashed lines are placed at
$V-I=0.95$ and $V-I=1.20$. \label{fig:gccol}} 
\vspace{-5 mm}
\end{figure}

Figure~\ref{fig:gccol} presents
the color distributions in each field. In every field, there is a peak
near $V-I=0.95$. The center field clearly shows a distinct peak near
$V-I=1.2$ as well. There are traces of this second peak in field~1 and
the in NGC~4486B field as well, but with smaller amplitude relative to
the blue peak. It appears that there are two populations of globular
clusters, the bluer of the two populations having a larger spatial
extent.
A Kolmogorov-Smirnov (K-S) test on the 3 fields with reasonably good
statistics rejects with a confidence of better then 97\% that the data
in each field arise from the same parent population, even if the
different fields are shifted to have the same mean color. The data are
consistent, however, with clusters in each of the fields arising from
a double Gaussian distribution, supporting the a model where the
central, red clusters are formed during mergers.

\vspace{-3 mm}
\section{Conclusions}
\vspace{-2 mm}
The surface brightness fluctuations and globular cluster luminosity
functions allow us to measure the distances to M87 and several of its
apparent neighbors, and determine which were true companions. Our
globular cluster catalogs provide valuable information on the
variation of the color distribution of globular clusters with the
distance from the center. However, our statistics were insufficient to
provide a more detailed understanding. In particular:
\begin{itemize}
\item M87, NGC~4486B, and NGC~4478 appear to be true companions, all
about 16 Mpc away. NGC~4476, on the other hand, is in the background,
at approximately 21 Mpc.
\item The K-S test on the globular cluster color distributions 
indicates that the samples detected at the 
different distances are unlikely to have arisen from the same
population.  We therefore conclude that the color distribution of the
globular cluster population in M87 varies with distance.
\item Even when the cluster populations are artificially shifted so
that they have the same mean color, the K-S test still indicates that
the different samples are unlikely to have arisen from the same
population. This casts doubt on models where all clusters have the
same origin.
\item The color distribution in the central field is significantly
better fit by two Gaussian distributions than one. Double peaked
distributions with the peak widths and the difference in peak
colors constrained to match the best fit double peaks of the center
fit the other fields well. This result is consistent with current
merger models. 
\end{itemize}

\vspace{-5 mm}
%

\end{document}